\documentclass[aoas,MSNbibl,nameyear,dvips]{arximspdf}
\usepackage{mathrsfs}
\usepackage{dcolumn}
\usepackage{graphicx}

\doi{10.1214/13-AOAS656} 
\volume{7}
\issue{3}
\pubyear{2013}
\firstpage{1593}
\lastpage{1611}

\makeatletter

\setattribute{copyright}{owner}{\textup{In the Public Domain}}

\newcolumntype{d}[1]{D{.}{.}{#1}}
\newcommand{\mbf}{\mathbf}
\newcommand{\bsy}{\bolds}

\newtheorem{thmm}{Result}
\newproclaim{defn}{Definition}
\makeatother

\begin{document}
\begin{frontmatter}

\title{Global space--time models for climate ensembles}
\runtitle{Global space--time models}
\begin{aug}
\author{\fnms{Stefano} \snm{Castruccio}\corref{}\thanksref{t1}\ead[label=e1]{castruccio@uchicago.edu}}
\and
\author{\fnms{Michael L.} \snm{Stein}\thanksref{t2}\ead[label=e2]{stein@galton.uchicago.edu}}
\thankstext{t1}{Supported in part by US NSF Grant 09-544(RDCEP) and by
STATMOS, an NSF funded
Network (NSF-DMS awards 1106862, 1106974 and 1107046).}
\thankstext{t2}{Supported by US NSF Grant 09-544(RDCEP).}
\runauthor{S. Castruccio and M. L. Stein}
\affiliation{University of Chicago}
\address{Department of Statistics\\
University of Chicago\\
5734 S. University Avenue\\
60637 Chicago, Illinois\\
USA\\
\printead{e1}\\
\phantom{E-mail:\ }\printead*{e2}}
\end{aug}

\received{\smonth{10} \syear{2012}}
\revised{\smonth{5} \syear{2013}}

%
\begin{abstract}
Global climate models aim to reproduce physical processes on a global
scale and predict quantities such as temperature given some forcing
inputs. We consider climate ensembles made of collections of such runs
with different initial conditions and forcing scenarios. The purpose of
this work is to show how the simulated temperatures in the ensemble can
be reproduced (emulated) with a global space/time statistical model
that addresses the issue of capturing nonstationarities in latitude
more effectively than current alternatives in the literature. The model
we propose leads to a computationally efficient estimation procedure
and, by exploiting the gridded geometry of the data, we can fit massive
data sets with millions of simulated data within a few hours. Given a
training set of runs, the model efficiently emulates temperature for
very different scenarios and therefore is an appealing tool for impact
assessment.
\end{abstract}

%

\end{frontmatter}

\section{Introduction}
There is a wide consensus among the scientific community that climate is
changing and this will bring significant imbalance to the present state
of the
system [IPCC AR4; \citet{me07}]. In order to assess the potential
impacts of
climate change both on the environment and human life, the geophysical
community is providing constantly growing ensembles of climate models that
include different scenarios of changing greenhouse gases (e.g., the CMIP5
archive [\citet{ta12}]). The advantages of a statistical analysis of climate
data lie in a framework that not only can provide insights about the ability
to reproduce the real climate, but also has crucial practical
advantages. If the
climate output can be reproduced efficiently with a simple statistical model
under some scenarios, then it is possible to predict how the output will
behave for a different scenario, both in terms of its mean and its covariance
structure. In other words, a statistical model can be used to fit some climate
model output under some scenarios and reproduce (\textit{emulate}) the
behavior of the climate model under a new forcing in much less time
than the
original computer run. This approach can provide policy makers with a powerful
tool for impact assessment. This work focuses on temperature at surface
for an
initial condition/scenario ensemble of a single General Circulation Model
(GCM), where the scenarios differ only in the trajectory of annual values
of CO$_2$ concentrations. The ultimate goal is to provide a statistical model
and to show how, given a small training set, it is possible to
reproduce the
computer output of other scenarios with only a small number of
parameters for
the mean and the covariance structure and a reasonable computational effort.
Annual averages of temperature at the pixel level are well-approximated
by a Gaussian distribution and we will use Gaussian process models throughout
this work.
A further simplification that we will examine and make use
of is that the covariance structure is independent of the scenario.

 To date, most work on climate model emulation has been done on
Regional Circulation Models (RCMs); see \citet{sa11a}, \citet{sa11b} and
especially \citet{gr11} for a statistical model of RCMs, and \citet{be12}
for a
model to adjust RCM output to real observations. Only a few studies
have been
conducted on statistical analysis of GCMs [see, e.g., \citet{ju08b} for
a statistical model for a multi-model GCM ensemble], and this is likely
due to the
dearth of literature regarding modeling data on the sphere$\times$time domain.
Recently, \citet{li11} introduced a Stochastic Partial Differential equation
approach to fit random fields. \citet{ju07,ju08} proposed a model
for processes on this domain based on taking derivatives of simpler
models and
\citet{ju11} extended it to the multivariate case. The latter approach
relies on embedding the sphere in $\mathbb{R}^3$, selecting an
isotropic model, and then applying partial derivatives to account for
anisotropies, directional effects and nonstationarities. This procedure
generates flexible models with explicit forms for the covariance
function, but
its coefficients are difficult to interpret so that it can be a
challenge to
specify forms of the model that would be appropriate in any particular
setting. The main contribution of our work is to introduce a spectral approach
in modeling GCM output that results in more interpretable coefficients,
improved fits and reduced computational cost for parameter estimation.

 Since a single climate run can contain several million simulated
values or even more for annual averages, care must be taken in fitting a
model. Current statistical methods to deal with massive space time data sets
often rely on different forms of a reduced rank approach, from fixed rank
kriging [\citet{cr08}] to predictive processes [\citet{ba08}]. Such
methods are
effective in fitting models in a feasible amount of time, but can
result in
loss of information and misfit [\citet{st08}]. In this work, the particular
geometry of the data set and the use of parallel computing achieve the
goal of
fitting the mean and covariance structure of a massive data set in a
few hours, by using a two-stage procedure that estimates some latitude
specific parameters separately for each latitude and then estimates a
few parameters describing dependence across parameters. The fitted
model, although not exactly the global maximizer of the likelihood
under the model, has a much higher likelihood than the maximized
likelihoods under current alternative models in the literature.

 Section~\ref{ensemble} presents the ensemble and explains how
the data are preprocessed. Section~\ref{statan} introduces a general
framework for the statistical analysis of this ensemble. Section~\ref{prosph} reviews the approach in \citet{ju07} for data on a spherical
domain. Section~\ref{spmodas} presents our spectral model, the spectral
decomposition of a class of spatial processes on a regular grid and
finally shows some results on the fit compared to current alternatives
in the literature. In order to emulate, Section~\ref{modmean} gives a
parametric model for the mean, extending the work in \citet{ca13}, and
shows an example of extrapolation for a different forcing scenario.
Section~\ref{conc} draws some conclusions.

\section{The ensemble}\label{ensemble}
\begin{figure}

\includegraphics{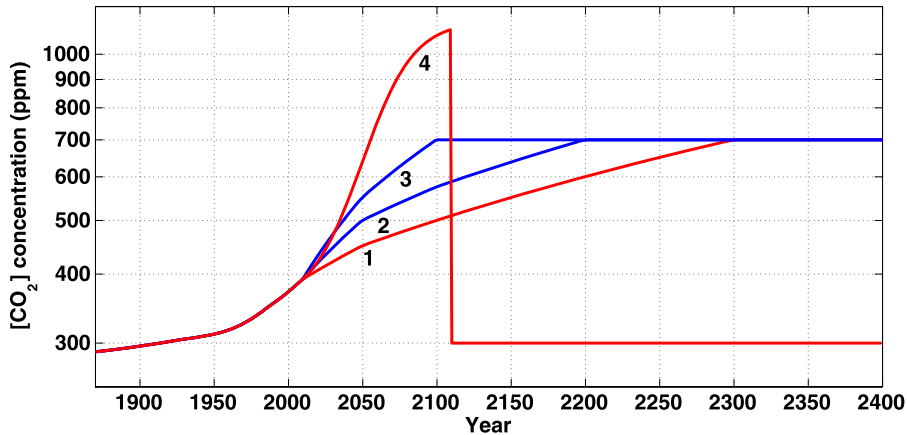}

\caption{Examples of CO$_2$ scenarios in the Climate Simulation
Library. We refer to these as
1: slow, 2: moderate, 3: high and 4: drop scenario. This work
mostly focuses on drop and slow
(in red), while moderate and high scenarios (in blue) will only be used
in the introduction of Section~\protect\ref{modmean}.}
\label{co2scen}
\end{figure}

We use the Climate Simulation Library of the Center for Robust Decision
Making on Climate and Energy Policy (RDCEP) consisting of model runs
made with the Community Climate System Model Version 3 [CCSM3; \citet
{YeagerJC06}, \citet{CollinsJC06}] at T31 resolution ($48\times96$ grid
points on a resolution of $\approx\!3.75^{\circ} \times3.75^{\circ}$).
The ensemble consists of multicentury model forecasts for a variety of
CO$_2$ trajectories (see Figure~\ref{co2scen} for some examples), with
all other greenhouse gas concentrations held constant at preindustrial
values. The relatively coarse spatial resolution allows generation of a
rich library for statistical analysis.

Our ensemble consists of multiple realizations for each scenario: $R=5$
initial conditions are sampled from the restart files of well spaced
out years of the NCAR b30.048 preindustrial control run [\citet
{CollinsJC06}]. For the purpose of this work these runs will be treated
as statistically independent, an assumption consistent with a
preliminary analysis of the data and physically realistic given the
extreme sensitivity to the initial conditions of the climate system.
From the CCSM3 output files we considered only the yearly average
temperature at surface, which we denote as $\mathbf{T}$. We also
removed the three northernmost and southernmost latitude bands to avoid
having to model the process in the narrow strips that form pixels near
the poles. A typical length of a single run is 500 years with $M=42$
latitudes and $N=96$ longitudes, so the data set has $42\times96\times
500\approx2$ million model simulated temperatures for each
realization.

\section{Statistical analysis of a climate ensemble}\label{statan}
In this section we explore some consequences of having independent
realizations of random fields in the ensemble. We consider runs with
the same
forcing and treat them as independent and identically distributed. We denote
with $L_m$ for $m=1,\ldots,M$ the latitude, with $\ell_n$ for
$n=1,\ldots,N$
the longitude, with $t_k$ for $k=1,\ldots,T$ the time.
The latitude bands do not need to be equally spaced in this framework.

 We will assume that the $r$th realization has distribution
%
%
\begin{equation}
\label{mod1} \mbf{T}_r=\bolds{\mu}+\bolds{
\varepsilon}_r,\qquad \bolds{\varepsilon}_r\sim
\mathcal{N}(\mathbf{0},\Sigma)
\end{equation}
for $r=1,\ldots, R$, where
\[
\mbf{T}_r=\bigl(\mbf{T}_r(L_1,
\ell_1,t_1),\ldots,\mbf{T}_r(L_M,
\ell _1,t_1),\mbf{T}_r(L_1,
\ell_2,t_1),\ldots,\mbf{T}_r(L_M,
\ell_N,t_T)\bigr)
\]
is the vector of temperatures, $\mathbb{E}(\mathbf{T}_r)=\bolds{\mu
}$ is
a mean, and $\bolds{\varepsilon}_r$ is the mean $\mathbf{0}$
stochastic component,
which is assumed to be normally distributed and with covariance matrix
$\Sigma$. We denote by $\bar{\mbf{T}}$ the mean across realizations and by
$S=TNMR$ the size of the data set made up of all realizations
of a scenario. If the data set consists of more than one realization,
we have that
$\mbf{T}_r-\mbf{T}_{r'}\sim\mathcal{N}(\mathbf{0},2\Sigma)$
for $r \neq r'$. Therefore, it is possible to
estimate the covariance structure without specifying a model for the mean.

\subsection{The restricted likelihood approach for the covariance
structure}\label{chreml}
Suppose now that the field has a parametrized covariance structure
$\Sigma=\Sigma(\bolds{\theta})$ that needs to be estimated. Also,
define $\mbf{T}= (\mbf{T}_1,\ldots,\mbf{T}_R )'$, $\mbf
{D}_r=\mbf{T}_r-\bar{\mbf{T}}$ and $\mbf{D}=(\mbf{D}_1,\ldots,\mbf
{D}_R)'$. By merging\vadjust{\goodbreak} all the different realizations, we can reformulate
(\ref{mod1}) as the following linear model:
%
%
\begin{equation}
\label{mod2} \mbf{T}= (\mbf{1}_R \otimes\mbf{I}_{TNM} ) (
\mbf {1}_R\otimes\bsy{\mu} )+\bolds{\varepsilon},\qquad \bolds {
\varepsilon}\sim\mathcal{N}\bigl(\mathbf{0},\mbf{I}_R \otimes\Sigma (
\bolds{\theta})\bigr),
\end{equation}
where $\mbf{I}_R$ is the $R \times R$ identity matrix and $\mbf{1}_R$
is a
column vector of length $R$ with all entries equal to 1. The design
matrix is
$\mbf{1}_R \otimes\mbf{I}_{TNM}$, each column allowing for different means
for every location in the grid and year, and $\bsy{\mu}$ the mean
parameter vector of length $TNM$.

 A natural way to estimate $\bolds{\theta}$ is by
restricted likelihood, and the following result gives an explicit formula.
%
\begin{thmm}\label{remlth}
The restricted loglikelihood for (\ref{mod2}) is
%
%
\begin{eqnarray}
\label{remlme} %
l(\bsy{\theta};\mbf{D}) & = & -
\frac{TNM(R-1)}{2}\log(2\pi)-\frac
{1}{2}(R-1)\log\bigl(\det\bigl(\Sigma(\bsy{
\theta})\bigr)\bigr)
\nonumber
\\[-8pt]
\\[-8pt]
\nonumber
& & -\frac{1}{2}TNM\log(R)-\frac{1}{2}\mbf{D}' \bigl(
\mbf{I}_R \otimes \Sigma(\bolds{\theta}) \bigr)^{-1}
\mbf{D}.
\end{eqnarray}
Also, the corresponding estimator for $\bsy{\mu}$ obtained by
generalized least squares is $\hat{\bsy{\mu}}=\bar{\mbf{T}}$.
\end{thmm}
The proof can be found in the supplementary material [\citet{supp}], along with further
theory on how the variogram can be estimated without bias in this context.
Result \ref{remlth} shows how adding independent realizations reduces to
summing $R$ quadratic forms for $\mbf{D}_r$ with the same matrix
$\Sigma^{-1}(\bolds{\theta})$, therefore, it does not require
storing matrices larger than in the case of a single realization.
Moreover, the
REML estimate of the mean vector $\bsy{\mu}$ does not depend on the covariance
structure and is just the sample average, which is expected since we assume
the realizations are independent and identically distributed.

\section{Processes on a spherical domain}\label{prosph}
Throughout this section we only consider the spatial part of the
process, so
we drop the time index. As the data have global coverage, a specific
statistical theory for random fields on a sphere is required. The
theory of
valid covariance functions on a sphere is different from that of the plane
[see \citet{gn12} for a complete discussion]. Furthermore, an isotropic process
on a sphere is not the natural choice in our case, as we expect temperature
fields to behave differently at different latitudes.
A more natural starting point is the following:
%
\begin{defn}
A Gaussian process $Z$ on a sphere is \textit{axially symmetric} [\citet
{jo63}] if it has mean only depending on latitude and
\[
\operatorname{cov}\bigl(Z(L_1,\ell_1),Z(L_2,
\ell_2)\bigr)=K(L_1,L_2,\ell_1-
\ell_2).
\]
Furthermore, the process is \textit{longitudinally reversible} [\citet
{st07}] if
\[
K(L_1,L_2,\ell_1-\ell_2)=K(L_1,L_2,
\ell_2-\ell_1).
\]
\end{defn}
In this work we only focus on axial symmetry although we believe that
such models are not fully adequate to describe surface temperatures. In
particular, accounting for land/sea differences may be one of the most
promising avenues for improving what we present here.

Axially symmetric models have seen a noticeable development recently.
\citet{ju07} proposed a constructive approach for generating such
processes, which results in an explicit form for the covariance function:

\begin{itemize}
\item define $k$ independent isotropic random fields $\tilde{Z}_j$,
$j=1,\ldots,k$ on $\mathbb{R}^3$,
\item consider its restriction on the unit sphere,
\item define
%
%
\begin{equation}
\label{pdeapp} Z(L,\ell):=\sum_{j=1}^k
\biggl(a_j(L)\frac{\partial}{\partial
L}+b_j(L)
\frac{\partial}{\partial\ell}+c_j(L) \biggr)\tilde{Z}_j(L,\ell),
\end{equation}
where
\begin{eqnarray*}
a_j(L) & = & \sum
_{i=1}^{n_{a_j}}a_{i,j} P^i\bigl(
\sin(L)\bigr),
\\
b_j(L) & = & \sum_{i=1}^{n_{b_j}}b_{i,j}
P^i\bigl(\sin(L)\bigr),
\\
c_j(L) & = & \sum_{i=1}^{n_{c_j}}c_{i,j}
P^i\bigl(\sin(L)\bigr)
 \end{eqnarray*}
and $P^i$ are Legendre polynomials of order $i$.
\end{itemize}

We will refer to this modeling framework as the partial derivative (PD)
approach. Using the PD approach guarantees that $Z$ is axially
symmetric; it can be extended to more general processes if $a_j$ and
$b_j$ depend on longitude and \citet{ju11} extends the approach to the
multivariate setting. Despite this flexibility, it has some
disadvantages. First, by starting out with models that must be valid in
$\mathbb{R}^3$, some possible models are lost, especially models with
substantial negative spatial correlation at some lags, which could
occur for quantities for which mass or energy are approximately
conserved over time. More importantly, the interpretation of the
coefficients $a_j$ and $b_j$ is not straightforward and limits the
flexibility of the model.

\section{Spectral modeling of axially symmetric processes}\label{spmodas}
We propose to represent the process in the spectral domain, and we show
how this results in a more flexible and interpretable model. We first
present the temporal part of the model, then define the model for a
single latitudinal band, a model for multiple latitudinal bands, the
structure of the spatial covariance matrix, and finally we compare this
model with the PD approach. We work under the assumptions of model (\ref
{mod1}). Sections~\ref{tempst}--\ref{multimod} describe our model and
summarize information about the parameter estimates. Section~\ref{spdec} shows how the axial symmetry of the spatial part of the model
can be exploited to speed up the calculations. Section~\ref{modcomp}
shows that our model yields much larger loglikelihoods than some PD models.

\subsection{The temporal structure}\label{tempst}
Define
$\tilde{\bolds{\varepsilon}}_t=(\varepsilon(L_1,\ell_1,t),\ldots
,\varepsilon(L_N,\ell_M,t))$
the vector of the variabilities at time $t$, and
$\mbf{D}_{t;r}=\mbf{T}_{t;r}-\bar{\mbf{T}}_t$ for $r=1,\ldots,R$ the
temperature difference. We assume that the vector-valued time
series $\tilde{\bolds{\varepsilon}}_t$ has the following structure:
%
%
\begin{eqnarray}\label{AR1}
\tilde{\bolds{\varepsilon}}_t &
=&\bolds{\Phi} \tilde {\bolds{\varepsilon}}_{t-1}+\bolds{
\eta}_t,\qquad \bolds {\eta}_t\sim\mathcal{N}(\mathbf{0},
\Sigma_s),
\nonumber
\\[-8pt]
\\[-8pt]
\nonumber
\bolds{\Phi} & =&\operatorname{diag}(\varphi_{Q(L_1,\ell_1)},\ldots,\varphi
_{Q(L_M,\ell_N)}),
\end{eqnarray}
where $Q(L,\ell)=1$ if pixel $(L,\ell)$ is land\setcounter{footnote}{2}\footnote{If the grid
point is on the boundary, we will consider it as land if its percentage
of land is greater than $50\%$.} and $Q(L,\ell)=0$ otherwise. In other
words, we are assuming a temporal AR(1) structure with different
correlation parameters, $\varphi_0$ and $\varphi_1$, depending on
whether the grid point is over land or ocean. Diagnostic plots (see the
supplementary material [\citet{supp}]) show that this structure is sufficient to
capture the temporal features of the data. We also assume $\tilde
{\bolds{\varepsilon}}_1\sim\mathcal{N}(\mathbf{0},\Sigma_s)$ so
the model is not exactly stationary in time, but given the weak
temporal correlation for annual temperatures, this simplification has
negligible impact on the model fit. For this section we define $\mbf
{H}_{t;r}=\mbf{D}_{t;r}-\bolds{\Phi}\mbf{D}_{t-1;r}$ and $\mbf
{H}_{r}=(\mbf{H}_{1;r},\ldots,\mbf{H}_{T;r})$.

\subsection{A model for a single latitudinal band}\label{singlemod}
Assume now that $\bolds{\eta}_t$ is axially symmetric, as
described in
Section~\ref{prosph}. If we consider a single latitudinal band, the covariance
$K_L$ is only a function of the longitudinal lag $\ell_n =2\pi n/N$,
$n=0,\ldots,N-1$ and is symmetric about $\pi$.
Therefore, we observe an evenly spaced stationary process on a
circle$\times$time domain and we define $f_L (c)=\sum_{n=0}^{N-1} e^{-i c
\ell_n}K_L(\ell_n)$ to be the spectral density on the circle at wavenumber
$c$. Since the grid we use here is the same grid on which a discretized
version of the partial differential equations underlying the GCM are solved,
it makes sense to work directly with this finite spectrum rather than
to model
a spectrum at all integer wavenumbers $c$.

 For observations on a line, a common spectral density is the
Mat\'ern:
%
%
\begin{equation}
\label{maternsp} f(\omega;\phi,\alpha,\nu)=\frac{\phi}{(\alpha^2+\omega^2)^{\nu
+1/2}},\qquad \omega\in
\mathbb{R}.
\end{equation}
We propose the following modification for $f_L$:
%
%
\begin{equation}
\label{maternci} f_L(c;\phi_L,\alpha_L,
\nu_L)=\frac{\phi_L}{ (\alpha_L^2+4\sin^2
({c}/{N}\pi ) )^{\nu_L+1/2}},\qquad c=0,\ldots,N-1.
\end{equation}
The parameters have similar interpretations as for the ordinary Mat\'ern
model, with $\alpha_L$ an inverse range parameter, $\nu_L$ controlling the
rate of decrease of the spectrum at large wavenumbers and thus the
``smoothness'' of the process (even though one cannot talk about
differentiability for a process on a discrete grid), and $\phi_L$ the overall
level of variation.

\begin{figure}

\includegraphics{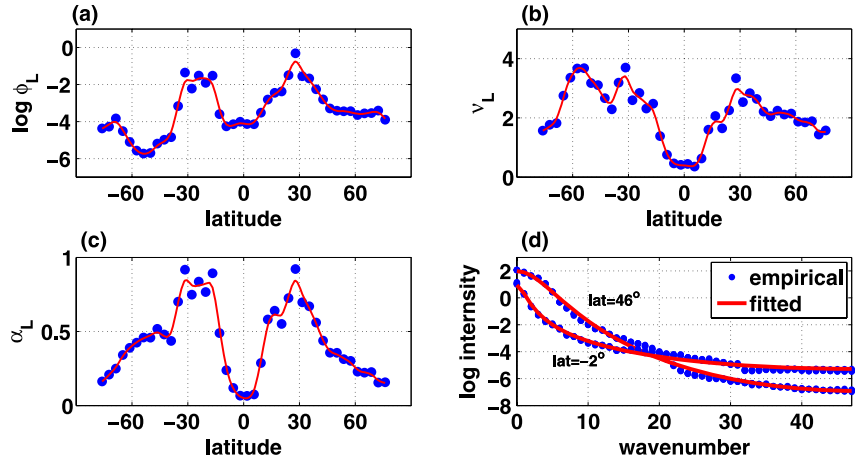}

\caption{(\textup{a)}: $\log(\hat{\phi}_L)$ for different latitudes. \textup{(b)}: $\hat
{\nu}_L$ vs latitude. \textup{(c)}: $\hat{\alpha}_L$ vs latitude. The blue dots
are the estimates and the solid red line is a fitted cubic smoothing
spline. \textup{(d)}: empirical and fitted log periodogram for two different
latitudinal bands, computed using the normalized differences $\sqrt {\frac{R}{R-1}}\frac{1}{T}\sum_{t=1}^T \hat{H}_{t;r}$ for $r=1,\ldots,5$.}
\label{singlelatplot}
\end{figure}

 A first analysis can be done by considering each band
separately from
the others. Figure~\ref{singlelatplot} shows the results for a training set
of five drop scenarios in which the parameters are estimated separately for
each latitude using REML.

In this way, one can visualize how $\hat{\phi}_L, \hat{\alpha}_L$ and
$\hat{\nu}_L$ in (\ref{maternci}) are changing across latitude [Figures~\ref{singlelatplot}(a)--(c)]. All the parameters display complex patterns;
it is
especially noticeable how $\hat{\alpha}_L$ and $\hat{\nu}_L$ have very
similar behaviors,
as they both show an increase at midlatitudes. The tropical behavior is very
different from all the other latitudinal bands, as estimates of both
parameters show a sharp drop in this region.

 Figure~\ref{singlelatplot}(d) shows an example of the periodogram fit
for two different bands: one near the equator and one at a northern midlatitude.
The spectrum near the equator
drops off faster at low wavenumbers, which is reflected in the
smaller value for $\hat{\alpha}_L$.
However, at high wavenumbers, the spectrum near the equator is flatter, which
is reflected in the smaller value for $\hat{\nu}_L$. The functional
form chosen is
flexible enough to capture the different behaviors across latitudes. The
supplementary material [\citet{supp}] provides a table of all the estimates with their
corresponding asymptotic standard deviations. The variability of these
estimates is extremely small, as we would expect from an analysis of
such a large
data set; therefore, the larger differences in patterns across
latitudes in Figures~\ref{singlelatplot}(a)--(c) are statistically
significant. In the
supplementary material [\citet{supp}], we further show how $\hat{\phi}_L, \hat{\alpha
}_L$ and
$\hat{\nu}_L$ do not substantially vary over time.

\subsection{A model for multiple latitudinal bands}\label{multimod}
To define a global model, we need to describe
the following:
\begin{itemize}
\item how $\phi_L$, $\alpha_L$ and $\nu_L$ are changing across
latitude,
\item how different latitude bands are correlated.
\end{itemize}
 The first point could be addressed with a parametric model of the
three parameters as a function of $L$, but, as shown in Figure~\ref{singlelatplot}, the pattern is complex and likely requires many
parameters to be adequately captured. Instead, we use the estimates obtained
from the analysis of single latitudinal bands. Computing estimates separately
for each band allows the estimates to be obtained in a parallel fashion on
multiple processors. In our analysis with $42$ latitudinal bands, the entire
procedure does not require more than 5 minutes on a small 4 node
cluster. In
contrast, the PD approach of \citet{ju08} does not lead to any obvious
algorithm to fit parameters that describe variation within latitude separately
for each latitude. With higher resolution model output, it might become more
important to parametrize how $\phi_L$, $\alpha_L$ and
$\nu_L$ change with latitude.

 As for the second point, we need to model
\[
f_{L_m,L_{m'}}(c)=\sum_{n=1}^N
e^{-i c \ell_n}K(L_m,L_{m'},\ell_n),\qquad c=0,
\ldots,N-1.
\]
More specifically, if we denote by $|\cdot|$ the modulus of a complex
number and by arg its argument, we need to specify coherence and the phase
\begin{eqnarray*}\rho_{L_m,L_{m'}}(c)& = & \frac{|f_{L_m,L_{m'}}(c)|}{\sqrt
{f_{L_m}(c)f_{L_{m'}}(c)}},
\\
\gamma_{L_m,L_{m'}}(c) & = & \operatorname{arg}\bigl(f_{L_m,L_{m'}}(c)
\bigr),
\end{eqnarray*}
where $f_{L_m}(c)$ and $f_{L_{m'}}(c)$ are defined in (\ref{maternci}).
A null phase results in a symmetric cross-covariance between bands,
which corresponds to a
symmetric circulant covariance matrix and therefore results in a
longitudinally reversible process (see Section~\ref{spdec}). The flexibility
of spectral methods allows one to account for longitudinal reversibility
independently from other features of the model. This is not possible in
the PD
approach: to have a longitudinally reversible process one needs $a_j(L)b_j(L)=0$
for $j=1,\ldots,k$ [\citet{ju08}] and it is not straightforward to determine
how such a constraint would impact other features of the model. Since
diagnostic plots (see supplementary material [\citet{supp}]) show that the phase is
small, we work under the assumption of longitudinal reversibility.

%
\begin{table}
\caption{Parameter estimates for the coherence (\protect\ref{cohmod})
in the
spectral model, for five drop scenarios. All estimates treat $\phi_L$,
$\alpha_L$ and $\nu_L$ for all $L$
as fixed at the values estimated in the first stage of
fitting}\label{cohest}
\begin{tabular*}{\textwidth}{@{\extracolsep{\fill}}lccc@{}}
\hline
\textbf{Parameter} & \textbf{Estimate} & \multicolumn{1}{c}{$\bolds{\mathrm{sd}\times10^{-4}}$} & \multicolumn{1}{c@{}}{$\bolds{95\%}$ \textbf{CI}} \\
\hline
$\xi$ & 0.9696 & 0.51 & (0.9695, 0.9697) \\
$\tau$ & 0.2080 & 2.0\phantom{0} & (0.2076, 0.2084) \\
$\varphi_{1}$ & 0.1010 & 3.4\phantom{0} & (0.1004, 0.1017)\\
$\varphi_{0}$ & 0.1141 & 3.5\phantom{0} & (0.1134, 0.1148)\\
\hline
\end{tabular*}
\end{table}

 We assume the following model for the coherence:
%
%
\begin{equation}
\label{cohmod} \rho_{L_m,L_{m'}}(c)= \biggl(\frac{\xi}{ (1+4\sin^2 (
{c}/{N}\pi ) )^{\tau}}
\biggr)^{|L_m-L_{m'}|},\qquad c=0,\ldots,N-1,
\end{equation}
where $\xi\in(0,1)$ and $\tau>0$. The proposed model has only 2 parameters:
$\xi$ controls the overall rate of decay of coherence across all wavenumbers
as the difference in latitude increases and $\tau$ describes how much faster
coherence decays at higher wavenumbers than at lower wavenumbers.
Note that we could allow $\tau<0$ as long as $\xi/5^\tau<1$ so that all
absolute coherences are bounded by 1 as they must be, but it would be very
unusual for a natural process to have stronger coherence across
latitudes at
higher wavenumbers. A more flexible form for (\ref{cohmod}) has been
considered but did not result in significant improvement of the fit
(see supplementary material [\citet{supp}]).

 Table~\ref{cohest} shows the estimated coefficients for the
five realizations of
the drop, together with their asymptotic standard deviations and $95\%$
confidence intervals by treating the previously estimated values of
$\phi_L$,
$\alpha_L$ and $\nu_L$ as known. We can see how all the estimates have very
small variability, which is expected since the data set is very large
($\approx\!
10.7$ million temperatures). The temporal structure is slightly
different for
land and ocean, as the latter tends to show a slightly stronger temporal
dependence. In the supplementary material [\citet{supp}] we further show how $\hat{\xi}$,
$\hat{\tau}$, $\hat{\varphi}_{1}$ and $\hat{\varphi}_{0}$ are not
dependent on
time and, therefore, the assumption of stationarity of the stochastic
term is
reasonable.

\subsection{Spectral decomposition of the covariance matrix for axially
symmetric processes}\label{spdec}
The evaluation of the likelihood in this setting is a challenging
problem, as
in general it requires evaluation of a quadratic form with an inverse
covariance matrix of size $TNM\times TNM$, and computation of a log
determinant. In this particular setting, the gridded geometry of the
data and
the axial symmetry allow for some degree of sparsity of $\Sigma_s$, the
covariance matrix for $\bolds{\eta}_t$, in the
spectral domain. Here we focus on describing matrix calculations for
$\Sigma_s$, which in turn with the AR(1) model in (\ref{AR1})
for $\bolds{\eta}_t$ allows for fast calculation of the restricted
likelihood.
For simplicity of notation, we drop
the time index $t$ throughout this subsection.

 In general, this problem requires $O((NM)^3)$ operations and the
storage of $NM(NM+1)/2$ distinct values using the Cholesky
decomposition. In
fact, the resolution of our model is sufficiently coarse that a general
Cholesky decomposition algorithm could be used here with some
difficulty. However, by exploiting
the structure of the covariance matrices, we can greatly speed up the
computation and reduce the memory requirement, which would be essential when
modeling higher resolution of GCM output.

 The regular lattice geometry for GCM output over the sphere, together
with the assumption that the model is axially symmetric, allows for
some exact
computations using spectral methods that drastically reduce the computational
time and the memory needed. This approach was first introduced by \citet{ju08}
for the analysis of Total Ozone Mapping Spectrometer (TOMS) Level 3 data,
which are post-processed data on a regular grid.

The key idea is that a stationary process of size $N$ on a circle
results in a (symmetric) circulant covariance matrix. The (real)
eigenvalues can be written in terms of the Fast Fourier Transform (FFT)
of the coefficients of the first row, which requires only $O(N \log
(N))$ operations, and the eigenvector matrix is simply the Discrete
Fourier Transform matrix [\citet{da79}, page~72].

 Over the sphere, $\bolds{\eta}_{L_m}=(\bolds{\eta
}(L_m,\ell_1),\ldots,\bolds{\eta}(L_m,\ell_N))$ is a process on a
circle for every $m=1,\ldots,M$. The $N \times N$ covariance matrix for
every $\bolds{\eta}_{L_m}$ is (symmetric) circulant and can
therefore be diagonalized via FFT. The $N\times N$ cross-covariance
matrix $\operatorname{cov}(\bolds{\eta}_{L_m},\bolds{\eta}_{L_{m'}})$
for $m\neq m'$ is circulant but not necessarily symmetric, as
\begin{eqnarray*} \operatorname{cov} \bigl(\bolds{
\eta}(L_m,\ell_n),\bolds{\eta }(L_{m'},
\ell_{n+s}) \bigr)&=& K(L_m,L_{m'},
\ell_n-\ell_{n+s})=K \biggl(L_m,L_{m'},2
\pi\frac{s}{N} \biggr)
\\
&\neq& K \biggl(L_m,L_{m'},-2\pi\frac{s}{N}
\biggr)\\
&=&K(L_m,L_{m'},\ell_n-\ell
_{n-(N-s)})
\\
&=&\operatorname{cov} \bigl(\bolds{\eta}(L_m,
\ell_n),\bolds{\eta }(L_{m'},\ell_{n-(N-s)})
\bigr).
\end{eqnarray*}
Therefore, the diagonalization via FFT results in complex
eigenvalues. It should be pointed out that the condition for the
cross-covariance matrix to be symmetric is that the process is
longitudinally reversible. If we call $\mathscr{F}$ the operation of
FFT, we know that the covariance matrix of $\{\mathscr{F}\bolds
{\eta}_{L_1},\ldots,\mathscr{F}\bolds{\eta}_{L_M}\}$ is a block
matrix with diagonal blocks, and if we rearrange rows and columns over
latitude, we have that $\mathscr{F}\bolds{\eta}$ is a block
diagonal matrix with $N$ blocks with each $M \times M$ block being an
Hermitian matrix. Therefore, the evaluation of the likelihood requires
$O (M^2 N\log N  )$ flops for the FFT and $O(M^3 N)$ for the
Cholesky decompositions of the $N$ blocks. In terms of memory, a
general axially symmetric process requires $M^2 N$ values to store,
while a longitudinally reversible process only requires $\frac
{M(M+1)}{2}N$ values.

%
\begin{table}[b]
\def\arraystretch{0.9}
\caption{Comparison between different models in terms of number of
parameters, computational time (hours) and restricted loglikelihood
(\protect\ref{remlme}). For the spectral model, the actual number of parameters for
the global maximization is reported in parentheses}\label{fitresults}
\begin{tabular*}{\textwidth}{@{\extracolsep{\fill}}ld{2.2}d{2.2}d{2.2}d{3.2}d{3.2}c@{}}
\hline
\textbf{Model} & \multicolumn{1}{c}{\textbf{\textsl{ind}}} & \multicolumn{1}{c}{\textbf{\textit{mat}}} &
\multicolumn{1}{c}{\textbf{\textit{h3}}} & \multicolumn{1}{c}{\textbf{\textit{h10}}} &
\multicolumn{1}{c}{\textbf{\textit{h3,2}}} & \multicolumn{1}{c@{}}{\textbf{\textit{sp}}}\\
\hline
\# param & 3 & 5 & 12 & 26 & 17 & 130 (4) \\
time (hours) & 0.1 & 7.3 & 60 & 379 & 880 & 1.8 \\
$\Delta$loglik/NMT(R-1) & -1.93 & -0.32 & -0.17 & -0.03 &
-0.16 & 0\\
\hline
\end{tabular*}
\end{table}

\subsection{Comparisons to other models}\label{modcomp}
We compare the spectral model\break (model~\textit{sp}) presented in the previous
two sections to several other models.
For all models, the temporal structure is given by the AR(1) model
(\ref{AR1}).
To model the spatial structure of the residual term $\bolds{\eta}_t$
in (\ref{AR1}), we consider the following possibilities:
a model with independent and identically distributed
components (model \textit{ind}), an
isotropic Mat\'ern model (model \textit{mat}) and the PD model with
Mat\'ern model for the underlying isotropic fields. Referring to equation
(\ref{pdeapp}), we consider the following settings for the PD models:
\begin{itemize}
\item Model \textit{h3}: $k=1$, $n_{a_1}=3$, $n_{b_1}=3$ and
$n_{c_1}=1$.
\item Model \textit{h10}: $k=1$, $n_{a_1}=10$, $n_{b_1}=10$ and
$n_{c_1}=1$.
\item Model \textit{h3,2}: $k=2$, $n_{a_1}=n_{a_2}=3$,
$n_{b_1}=n_{b_2}=3$, $n_{c_1}=1$ and $n_{c_2}=0$.
\end{itemize}

We use 5 realizations of a drop scenario, for a total of approximately 10.7
million GCM temperatures, and we compare the results in terms of the
restricted loglikelihood (\ref{remlme}). For such massive data sets,
we found it helpful to normalize differences in (restricted)
loglikelihoods by the number of contrasts $NMT(R-1)$.
We will write $\Delta$loglik to generically mean a difference in
loglikelihoods.
For the \textit{sp} model, we first
maximize the likelihood for the single band parameters in parallel and
then we
maximize the likelihood for $\xi, \tau, \varphi_0$ and $\varphi_1$
conditional on the values of
$\hat{\phi}_L$, $\hat{\alpha}_L$ and $\hat{\nu}_L$.
All the other models are maximized over the full parameter space.
The results are reported in Table~\ref{fitresults},\vadjust{\goodbreak} where the number of
parameters and the difference in loglikelihood (normalized with respect
to the size of the data set) is shown and compared. The
model \textit{ind} is clearly not adequate, and the isotropic
Mat\'ern results are a noticeable improvement. Model \textit{h3} gives better
results, therefore underlying the need for an anisotropic model. Model
\textit{h3,2} and especially \textit{h10} result in better likelihoods but
the number of parameters is very large, and the estimation requires several
weeks. Model \textit{sp} outperforms all the previous models and even
if the
actual number of parameters is $4+42\times3=130$, the maximization is done
only with the 4 parameters in Table~\ref{cohest} and requires only 1.8 hours.
The precomputation of the parameters for the latitudinal bands (a procedure
that, as mentioned in Section~\ref{singlemod}, requires a few minutes using
multiple processors on a cluster) plays a crucial role in this model,
as it
adds flexibility and allows for a maximization of a conditional loglikelihood
with respect to only a few parameters.

Although the model presented here already provides a substantially
better fit
than even the best PD model with much less computation, one might
wonder if
maximizing the restricted loglikelihood over all 130 parameters would
lead to
a model with a much better fit. We ran a full parameter search over all 130
parameters using \verb|fminsearch| in MATLAB, which resulted in an
improvement of $\approx\!0.008 \Delta$loglik$/NMT(R-1)$ after
approximately 1670 hours of computation.
Our goal here is to find the best-fitting model to the data that we can
for a given computational effort and it is clear that the \textit{sp}
model with parameters estimated by our proposed two-stage procedure
dominates the PD models with parameters estimated by REML in this application.

\begin{figure}

\includegraphics{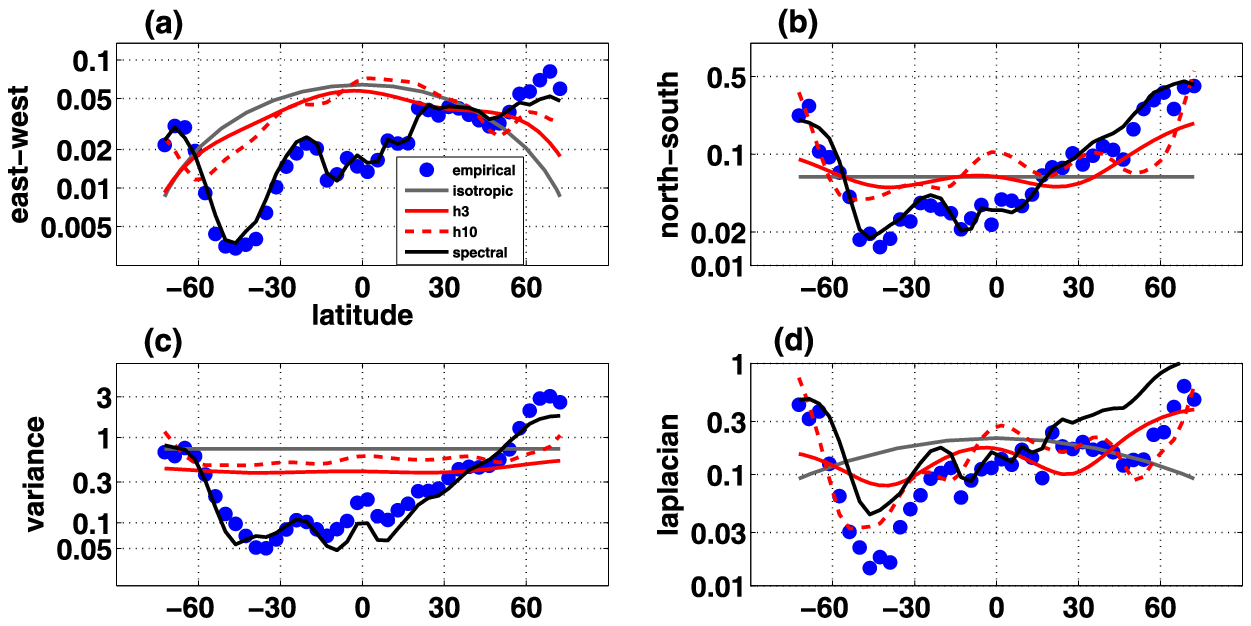}

\caption{\textup{(a)}: estimated and fitted $\operatorname{var}(\mbf{H}(L,\ell)-\mbf{H}(L,\ell
-\Delta\ell))$. \textup{(b)}: $\operatorname{var}(\mbf{H}(L_m,\ell)-\break \mbf{H}(L_{m-1},\ell))$.
\textup{(c)}: $\operatorname{var}(\mbf{H}(L,\cdot))$. (d): $\operatorname{var}(4\mbf{H}(L_m,\ell)-\mbf
{H}(L_m,\ell-\Delta\ell)-\mbf{H}(L_m,\ell+\Delta\ell)-\break \mbf
{H}(L_{m-1},\ell)-\mbf{H}(L_{m+1},\ell))$. The vertical axis is plotted
on a log scale. The details about computing the empirical estimates are
in the supplementary material [\citet{supp}].}
\label{diagnmod}
\end{figure}

To have a better understanding of how the model is able to capture the local
spatial dependence of the data, it is useful to show how variances of
spatial contrasts are reproduced by the model [see, e.g., \citet{st04}]. Figure~\ref{diagnmod} shows a comparison between models \textit{mat}, \textit{h3},
\textit{h10} and \textit{sp} in terms of their ability to reproduce the
variances of some contrasts of $\mbf{H}_r$\footnote{The index for $\mbf
{H}_r$ will be dropped,
as the distribution of the contrasts is independent of the realization; note
that $\mbf{H}_r$ depends on the estimates of the AR(1) parameters $\phi_0$
and $\phi_1$, but the values of these parameters vary so little across models
that the visual impression of Figure~\ref{diagnmod} is unaffected by which
estimates are used.} (the
details about how the empirical estimates are computed are found in the
supplementary material [\citet{supp}]). Figures~\ref{diagnmod}(a)--(b) represent the east--west
contrast $\mbf{H}(L_m,\ell_n)-\mbf{H}(L_m,\ell_{n-1})$ and the north--south
contrast $\mbf{H}(L_m,\ell_n)-\mbf{H}(L_{m-1},\ell_n)$.
In both cases the spectral
model is able to reproduce the patterns of the empirical contrast, therefore
showing an overall good fit of both the single band spectral model (east--west)
and the coherence (north--south). The isotropic model shows a pattern in the
east--west contrast that is only due to the geometry of the sphere: points
closer to the poles are physically closer for the same longitude spacing,
therefore resulting in smaller variances of the contrasts. Model \textit{h3}
instead is able to capture some of the features of the data, especially for
the north--south contrast, but is overall too smooth and would require more
flexibility. Model \textit{h10} shows a decent fit in the north--south
contrast, but the east--west contrast is significantly misfitted. Figures~\ref{diagnmod}(c)--(d) represent the variance across latitudes and the Laplacian.
The spectral model can reproduce most of the trend for the variance, therefore
proving to be flexible enough to capture the low wavenumbers behavior. The
Laplacian is overestimated for the northern hemisphere, a sign of a
lack of
fit for the coherence between multiple bands that could be fixed by allowing
nonstationarity across latitudes in (\ref{cohmod}), although at the
cost of a
substantially more complex model. The model \textit{h10} shows a somewhat
overall better fit for this contrast.

\begin{figure}

\includegraphics{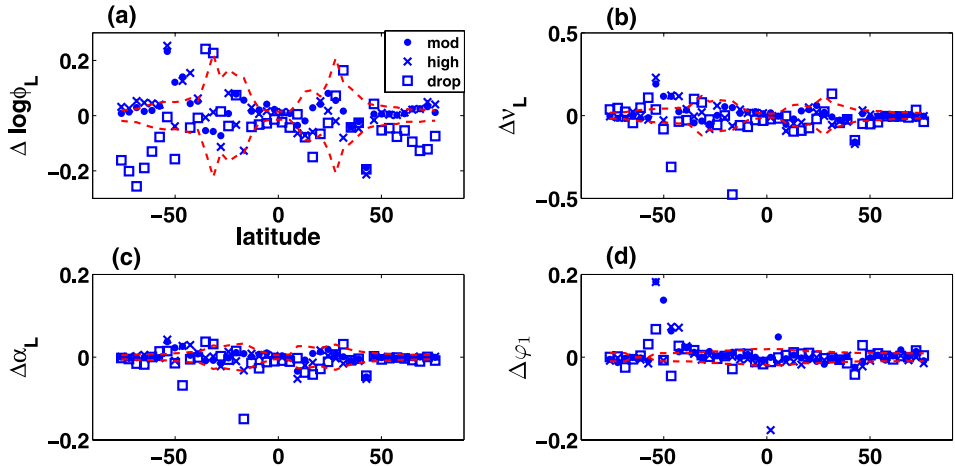}

\caption{Differences of parameter estimates for different scenarios
($R=5$ realizations each) for \textup{(a)}: $\log\phi_L$, \textup{(b)}: $\nu_L$,
\textup{(c)}:
$\alpha_L$ and \textup{(d)}: $\varphi_1$. The reference is the slow scenario. The
dashed red lines represent the $95\%$ Bonferroni confidence bands
around 0. The estimates are computed as in Figure~\protect\ref{singlelatplot}.}
\label{errorscenario}
\end{figure}

\section{Emulation}\label{modmean}
Throughout this section, we drop the index $r$ denoting the realization to
simplify the notation. The model we have presented can reproduce
efficiently the
covariance structure for a single scenario, but in order to emulate
temperature for a
different forcing, we need to determine how the mean and the covariance
structure are changing across different scenarios. Figure~\ref{errorscenario}
addresses the latter issue. The plots represent the change in parameter value
for the single latitudinal band features across four scenarios
indicated in
Figure~\ref{co2scen}; $\varphi_{1}$ is not included but shows similar
patterns. For all of them, an analysis of $R=5$ different realizations is
performed and we show the differences with respect to the slow scenario,
together with the $95\%$ Bonferroni confidence bands around 0. Since
all the
standard deviations are similar across scenarios, we choose to plot the
differences of parameter estimates between the slow and the drop
scenario. The
differences between the slow and drop scenarios are significant for
$\log\phi_L$ at higher latitudes, as shown in Figure~\ref{errorscenario}(a),
but all the other parameters and scenarios are largely within the confidence
bands. Therefore, it seems reasonable to assume that
differences in the covariance structure across scenarios are modest.
In the supplementary
material [\citet{supp}], a similar diagnostic is carried out for the coherence parameters.
Therefore, only a parametrization of the mean is necessary to describe the
trend of the data, train the model with some scenarios, and then
predict the
temperature for an unknown scenario. Our approach to emulation of the
mean is
based on the method of \citet{ca13}, where a simple model was proposed
and the analysis
was performed independently for 47 regions without accounting for spatial
dependence.

\subsection{A model for the mean}
Before proceeding with this analysis, we standardize to account for the
different variability across different grid points. This simple
adjustment was not done in the previous sections and accounts for some
of the nonstationarity in longitude that our model is not able to
capture. In this section we consider the output of a control run with
constant CO$_2$ concentration, compute its average $\bar{\mathbf{T}}_c$
and its standard deviation $\operatorname{sd}(\mathbf{T}_c)$ over time for
every grid point, and then for every temperature vector $\mbf{T}$ in
the training set normalize to $\mathbf{T}^*:=(\mathbf{T}-\bar{\mathbf
{T}}_c)/\operatorname{sd}(\mathbf{T}_c)$.

 With reference to (\ref{mod1}), the goal of this section is to
give a (scenario dependent) parametric model for $\bolds{\mu}$ in
order to extrapolate the temperature value for another given scenario.
The model we use is the following:
%
%
\begin{eqnarray}
 \mbf{T}^*(L,\ell,t) & = & \beta_{0,(L,\ell)}+
\beta_{1,(L,\ell)} \biggl(\frac{1}{2}\log[\operatorname{CO}_2](t)+
\frac{1}{2}\log[\operatorname{CO}_2](t-1) \biggr)
\nonumber\\
& &{} +\beta_{2,c}\sum_{i=2}^{+\infty}w(i-2)
\log[\operatorname {CO}_2](t-i)+\varepsilon(L,\ell,t),
\\
w(i) & = & \lambda^i (1-\lambda),\nonumber
\end{eqnarray}
where $\varepsilon$ is modeled as in Section~\ref{spmodas} and $c$ indexes
the 47 predefined regions shown in the supplementary materials [\citet{supp}].

The mean has three components:
\begin{itemize}
\item an intercept $\beta_0$, different for every grid point,
\item a short term effect $\beta_1$, different for every grid point,
\item a long term effect $\beta_2$, different for the 47 regions.
\end{itemize}
The last term accounts for the long term contribution of the forcing
via the
weights $w(i-2)$, which are expected to be decreasing as the time lag $i-2$
increases. We model this decrease exponentially with decay rate $\lambda$
identical for every grid point. The linear parameters can be profiled
but the
estimation of the parameters describing the behavior of
$\bolds{\varepsilon}$ and of $\lambda$ requires the numerical
maximization of the likelihood. The evaluation of this likelihood for a single
run requires approximately 8 minutes, so to make the computation faster and
reduce the optimization to $\lambda$, we plugged in the estimated
spatiotemporal structure of $\bolds{\varepsilon}$ obtained via REML
using the same procedure as in the previous section.

 We estimated the stochastic structure with $R=5$ realizations
of the
drop scenario, obtained $\lambda$ based on a single drop scenario
(estimation of $\lambda$ for all scenarios was not computationally
feasible), and finally
emulate for the slow scenario. We choose the drop scenario for the
training set
to show the results for a severe extrapolation and because the trend is more
evident under a forcing with an abrupt change, making the estimation of the
coefficients more stable. In order to reproduce the mean climate for less
drastic scenarios such as the Representative Concentration Pathways in the
CMIP5 archive [\citet{vv11}], a simpler form of the mean function is likely
preferable.

\begin{figure}

\includegraphics{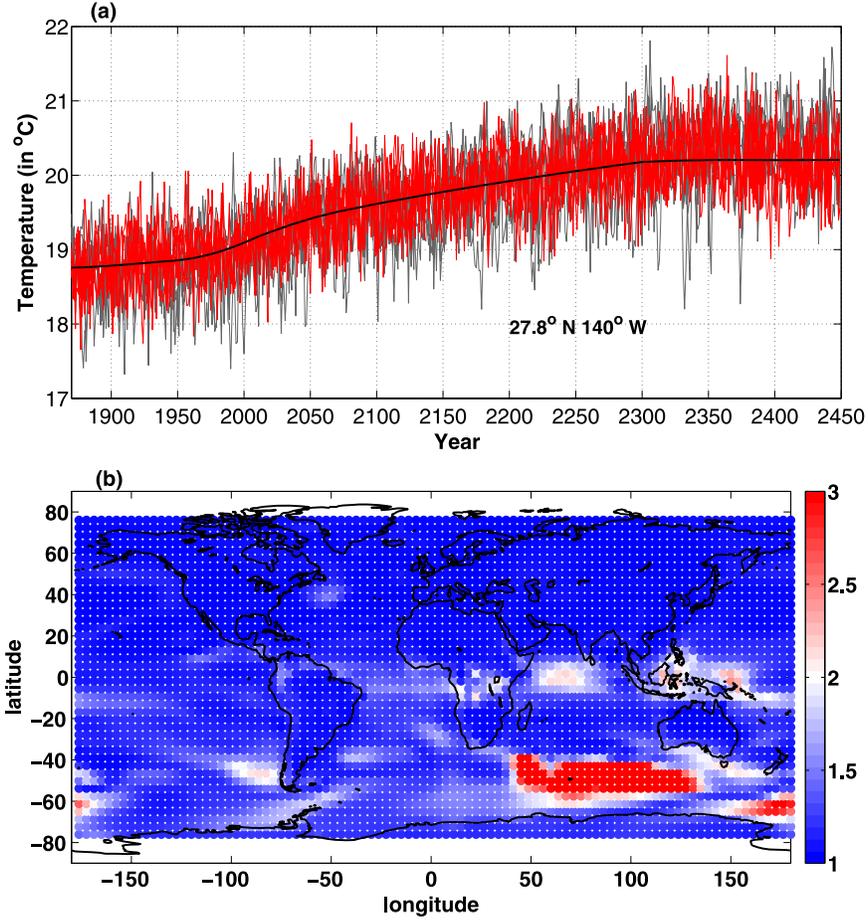}

\caption{Training set: one drop realization. Prediction set: a slow scenario.
\textup{(a)}: an example of a conditional simulation. The solid black line represents
the fitted (emulated) value, the gray lines represent the actual realizations
from the GCM and the red lines are the conditional simulations. \textup{(b)}:~global
plot of the lack of fit index $I$, which is a measure of goodness of
fit for
emulation of the mean.
The upper bound at $I=3$ is only for visualization purposes as the
fit index can be as large as 18.9. There are 74 out of 4032 points with
$I>3$.}
\label{globalan}
\end{figure}

 The estimated value for $\lambda$ is $0.95$ $(0.0014)$ and once this
parameter is estimated, each conditional simulation takes only a few seconds.
Given the very large number of temperatures and the small variability
of the
estimates, we choose not to account for parameter uncertainty in the
simulations. The top part of Figure~\ref{globalan} gives an example of
emulation of the mean
and conditional simulation\vadjust{\goodbreak} of the slow scenario for a grid point in
the middle of the Pacific Ocean. To assess the fit, for every grid
point we
use, as in \citet{ca13},
the following simple lack of fit index (the indication of latitude and
longitude has been removed for simplicity):
%
%
\begin{equation}
\label{I1} I=\frac{\sum_{r=1}^R \sum_{t=1}^T (\mbf{T}_{r}(t)-\hat{\mbf
{T}}(t))^2}{({R}/{(R-1)}) \sum_{r=1}^R \sum_{t=1}^T (\mbf{T}_{r}(t)-\bar
{\mbf{T}}(t))^2},
\end{equation}
where $\hat{\mbf{T}}$ is the fitted value and $\bar{\mbf{T}}$ is the
average across realizations. This index measures how close the fitted
value (emulated mean) is to the average of the realizations. The
smaller $I$ is the better, but since the expected value of the
numerator cannot be less than that of the denominator, values of $I$
near 1 indicate an excellent fit.

Figure~\ref{globalan}(a) shows an example of conditional simulation for a point
in the Pacific Ocean. The mean and variation are similar to the original
simulations, but there is noticeable underestimation of extreme events,
especially cold extremes. This lack of ability of the statistical model to
reproduce such features of the climate poses some problems and can
limit the
extent of the use of this model on impact assessment. A possible
direction to
address this issue is to fit the data with a model more general than an
axially symmetric, but this will require further advances in modeling, and
substantial computational resources. Another approach is to develop
direct methods to fitting and emulating extremes along the lines of
\citet{ma10}.

 Figure~\ref{globalan}(b) shows $I$ for the slow scenario over all
the regions. It is evident that the emulation does poorly in the area
near the Southern Ocean stretching from the southern tip of South
Africa to near Tasmania. We speculate that the sea ice albedo effect
may create strong nonlinearities in this area, but further
investigations are needed. Also, the fit is not fully adequate in the
equatorial regions, even though the misfit is not as strong as in the
Southern Ocean. The details about the algorithm are provided in the
supplementary material [\citet{supp}], and the file called climate\_movie shows a
movie of a conditional simulation in terms of anomalies from
preindustrial conditions.

\section{Conclusions}\label{conc}
Spectral modeling is a natural choice for gridded data on sphere$\times
$time, and we have shown how it outperforms the current alternatives in
the literature in terms of simplicity, flexibility and computational
requirements. Although this work has focused on temperature, a similar
approach can be extended to other climate variables such as
precipitation [see \citet{ca13} for an example of mean emulation] and
possibly different time scales, as long as the normality hypothesis is
tenable. On a single latitudinal band, our model assumes independence
of the spectral process across wavenumbers, but this assumption can in
principle be relaxed to account for more complex nonstationarities. We
have also shown an example of fit for climate model output of
challenging size, and we have carried out the analysis without
appealing to reduced rank representations of the process. The specific
geometry of the climate output has played an important role, but
distributing some parts of the algorithm across different processors
has also contributed to reduce the computational burden, and further
work is needed to understand how parallel computation can be helpful in
fitting massive data sets.

\section*{Acknowledgments}
The authors thank Elisabeth Moyer and David McInerney at the Department
of Geophysical Sciences at the University of Chicago for providing the
ensemble data and for the useful discussions. This work is part of the
RDCEP effort to improve the understanding of computational models
needed to evaluate climate and energy policies.

\begin{supplement}[id=suppA]
\stitle{Supplementary material}
\slink[doi]{10.1214/13-AOAS656SUPP} 
\sdatatype{.pdf}
\sfilename{aoas656\_supp.pdf}
\sdescription{Further technical details and theoretical results can be found in the
online supplementary material.}
\end{supplement}

%

%

\printaddresses

\end{document}